\documentclass[showpacs,amssymb,twocolumn,aps,nofootinbib]{revtex4}
\usepackage{amsmath}
\usepackage{amstext}
\usepackage{amsopn}
\usepackage{amsfonts}
\usepackage{amssymb}
\usepackage{bbm}
\usepackage{accents}
\usepackage{empheq}
\usepackage{graphicx}
\usepackage{epsf}
\usepackage{graphics}
\usepackage[latin1]{inputenc}
\def\sc{\scriptstyle}

\def\sl{\!\!\!\slash}

\begin{document}

\title{The pole of the fermion propagator in a general class of gauges}

\author{Ashok K. Das$^{a,b}$ and J. Frenkel$^{c}$ \footnote{$\ $ e-mail: das@pas.rochester.edu,  jfrenkel@fma.if.usp.br}}
\affiliation{$^a$ Department of Physics and Astronomy, University of Rochester, Rochester, NY 14627-0171, USA}
\affiliation{$^b$ Saha Institute of Nuclear Physics, 1/AF Bidhannagar, Calcutta 700064, INDIA}
\affiliation{$^{c}$ Instituto de Física, Universidade de São Paulo, 05508-090, São Paulo, SP, BRAZIL}

\begin{abstract}
We study the behavior of the pole of the fermion propagator, in QED in $n$-dimensions, in a general class of gauges which interpolate between the covariant, the axial and the Coulomb gauges. We use Nielsen identities, following from the BRST invariance of the theory, to determine the gauge variation of the fermion two point function in this general class of gauges. This allows us to show directly  and in a simple manner, to all orders in perturbation theory, that in the absence of infrared divergences and mass shell singularities, the fermion pole mass is gauge independent.
\end{abstract}


\maketitle

In a relativistic quantum field theory and, in particular, in an interacting theory of fermions, the physical mass of the fermion is determined from the pole of its complete propagator and is calculated, in perturbation theory, order by order. When the fermion is interacting with gauge particles, the fermion two point function and, therefore, its propagator become gauge dependent. There are two sources of gauge dependence that arise in this case. First, within any class of gauges (say, covariant or axial or Coulomb), the two point function and the propagator become functions of the gauge fixing parameter. Second, the forms of the two point function and the propagator are different in different classes of gauges. Therefore, it is not clear {\em a priori} that the pole of the propagator will be gauge independent (in both senses). On the other hand, if the pole of the propagator is to correspond to the physical mass of the particle, it must be gauge independent in the sense that it should not only be independent of the gauge fixing parameter, it must also have the same value in different classes of gauges.

The question of the gauge parameter independence of the pole of the fermion propagator has been  studied in the literature mainly in the class of covariant gauges \cite{atkinson,tarrach,johnston,reinders,gray,brown,brecken,kronfeld}.In particular, in covariant gauges the gauge parameter independence for the fermion pole mass has been addressed in \cite{johnston,brecken} using the Nielsen identity \cite{nielsen, das,kobes} which determines the gauge variation of the fermion two point function. However, a systematic study of this question to all orders in non-covariant (axial, Coulomb) gauges \cite{kummer, frenkel, leibbrandt, fradkin,bernstein,andrasi} is lacking so far. Furthermore, to the best of our knowledge, it has not been demonstrated explicitly that the pole of the fermion propagator occurs at the same value in different classes of gauges which we would expect on physical grounds. In this analysis, we choose a general class of gauges \cite{taylor} which interpolate between the covariant, the axial and the Coulomb gauges to study this question. With such a choice of gauge, we show in a unified manner that in a theory without any infrared divergences and mass shell singularities, the pole of the fermion propagator is not only independent of the gauge fixing parameters, but also has the same value in the covariant, the axial and the Coulomb gauges. This simple and direct proof of gauge independence involves three basic ingredients, namely, choosing a general class of gauges which interpolate between different gauges, the Nielsen identity  for the gauge variation of the fermion two point function and the relation of this function to the denominator of the propagator. In this letter we only sketch the essential steps in the proof of gauge independence, leaving the details of the derivations  to a separate publication \cite{new}.

Let us recapitulate briefly what is already known in the literature in covariant and axial gauges. In the covariant gauge, since the momentum of the particle, $p_{\mu}$, is the only Lorentz four vector available, one can parameterize the fermion self-energy (to all orders) as
\begin{equation}
\Sigma^{\rm (c)} (p) = m A  + B p\sl,\label{1}
\end{equation}
where $m$ denotes the tree level fermion mass,  $A, B$ are dimensionless coefficients, functions of  $(p^{2},m)$ as well as the gauge fixing parameter and the superscript (c) stands for the covariant gauge. If the propagator has a pole at $p^{2} = M^{2}_{\rm (c)}$, then the fermion dynamical equation can be written as
\begin{align}
S_{\rm (c)}^{-1} (p) u (p)\Big| & = (p\sl - m - \Sigma^{\rm (c)}) u (p)\Big|\notag\\
& = (p\sl - M_{\rm (c)})u(p)\Big| = 0.\label{1a}
\end{align}
Here the restriction $|$ implies that these quantities are evaluated at the pole of the propagator $p^{2} = M^{2}_{\rm (c)}$. It follows from \eqref{1a} that 
\begin{equation}
\overline{u}\, S_{\rm(c)}^{-1} (p)\, u\Big| = \overline{u}\, (p\sl - m - \Sigma^{\rm (c)}(p))\,u\Big| = 0,\label{2}
\end{equation}
which allows us to identify
\begin{equation}
M_{\rm (c)} = m + \overline{u}\, \Sigma^{\rm (c)} (p)\, u\Big|.\label{3}
\end{equation}
Consequently, studying the gauge fixing parameter dependence of $M_{\rm (c)}$ is equivalent to studying the gauge parameter dependence of $\overline{u}\, \Sigma^{\rm (c)} (p)\, u$ at the pole of the fermion propagator. 

In non-covariant gauges, however, in addition to the four momentum of the particle, we have a direction vector $n^{\mu}$ (we choose $n^{2}\neq 0$ for simplicity). As a result, the structure of the fermion self-energy becomes a bit more involved and is conventionally parameterized in the form \cite{pagels, konetschny}
\begin{equation}
\Sigma^{\rm (nc)} (p) = m A + B p\sl + C p\sl_{\sc L} + \frac{m D}{p^{2}_{\sc L}}\left(p\sl_{\sc L} p\sl - p\sl p\sl_{\sc L}\right),\label{4}
\end{equation}
where $p^{\mu}_{\sc L}$ denotes the component of the four momentum $p^{\mu}$ along the given direction $n^{\mu}$, namely,
\begin{equation}
p^{\mu}_{\sc L} = \frac{(n\cdot p)}{n^{2}}\,n^{\mu},\label{5}
\end{equation}
and the coefficients $A, B, C$ and $D$ are, in general, dimensionless functions of $(p^{2},p^{2}_{\sc L}, n^{2}, m)$. It is clear that not having an additional direction $n^{\mu}$ (or $p^{\mu}_{\sc L}$) is equivalent to having the coefficients $C=D=0$ in which case \eqref{4} reduces to \eqref{1}. The dependence on this additional direction makes the extraction of the physical mass more complicated. Conventionally, one assumes that this mass, $p^{2} = \widetilde{M}^{2}$, can be obtained, as in covariant gauges (see \eqref{2}), from the relation
\begin{align}
\overline{u} S^{-1}_{\rm (nc)} (p) u\Big| & = \overline{u}(p\sl - m - \Sigma^{(nc)}) u\Big|\notag\\
& = \overline{u} (p\sl -\widetilde{M})u\Big|= 0.\label{5a}
\end{align}
This would then determine, as in \eqref{3}, that the mass $\widetilde{M}$ is given by 
\begin{equation}
\widetilde{M} = m + \overline{u}\,\Sigma^{\rm (nc)}\, u\Big|_{p^{2}=\widetilde{M}^{2}}.\label{6}
\end{equation}
At one loop, from the simplicity of the integrand for the fermion self-energy, it is easily seen \cite{konetschny} that $\widetilde{M}$ is gauge independent (independent of $n^{\mu}$).  Based on this calculation, it has been  assumed that the mass $\widetilde{M}$ determined from \eqref{6} is gauge independent to all orders as in covariant gauges. 

\begin{widetext}
\begin{figure}[ht!]
\begin{center}
\includegraphics[scale=.8]{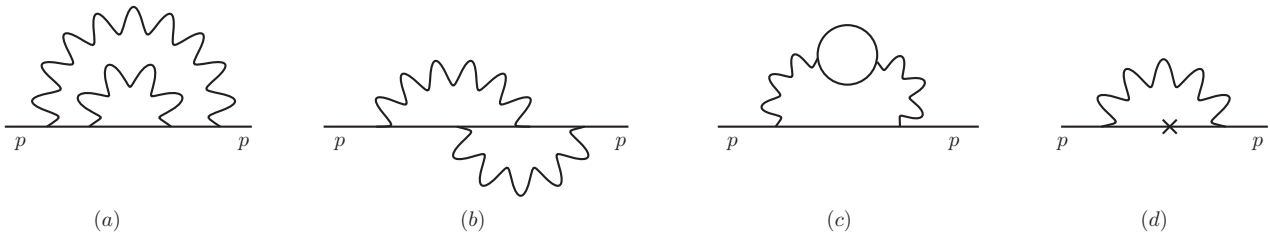}
\end{center}
\caption{Feynman diagrams for the fermion self-energy in QED at two loops.}
\label{fig1}
\end{figure}
\end{widetext}

However, an explicit calculation of the fermion self-energy at two loops shows that $\overline{u}\,\Sigma^{\rm (nc)}\,u$ is manifestly gauge dependent at $p^{2} = \widetilde{M}^{2}$. We will explain this calculation briefly here leaving more details to \cite{new}. The mass $\widetilde{M}$ in \eqref{6} at two loops can be determined explicitly from the Feynman diagrams in Fig. \ref{fig1}
Here the diagrams $(a), (b)$ and $(c)$ denote the standard two loop diagrams for the self-energy while $(d)$ represents the contribution coming from the one loop mass correction. In the generalized axial gauge, the photon propagator in QED has the form
\begin{widetext}
\begin{equation}
D_{\mu\nu} (p) = - \frac{1}{p^{2}}\left[\eta_{\mu\nu} - \frac{n_{\mu}p_{\nu}+n_{\nu}p_{\mu}}{n\cdot p} + \frac{p_{\mu}p_{\nu} n^{2}}{(n\cdot p)^{2}}\left(1 + \frac{n^{2}p^{2}}{\beta^{2} (n\cdot p)^{2}}\right)\right],\label{6a}
\end{equation}
\end{widetext}
where $\beta$ denotes the gauge fixing parameter. Since the fermion loop is transverse, all the momentum $p^{\mu}$ dependent terms coming from the photon propagator \eqref{6a} vanish in diagram $(c)$ and it is independent of $n^{\mu}$ and $\beta$. The other three diagrams are independent of the parameter $\beta$ (but do depend on $n^{\mu}$) because of the principal value prescription used in axial gauges. There are several  cancellation of the $n^{\mu}$ dependent terms in the combination of the graphs $(a), (b)$ and $(d)$ and the final form of $\widetilde{M}$ at two loops has the form
\begin{widetext}
\begin{equation}
\widetilde{M} = M^{(Feynman)} + \frac{e^{4}}{(2\pi)^{2n}}\,\overline{u} (p) \int \frac{d^{n}q_{1}}{q_{1}^{2} (n\cdot q_{1})}\, n\sl \frac{1}{p\sl + q\sl_{1} - m}\, (p\sl - m) \int \frac{d^{n}q_{2}}{q_{2}^{2} (n\cdot q_{2})}\,\frac{1}{p\sl + q\sl _{2} - m}\, n\sl\, u (p)\Big|_{p^{2}=m^{2}},\label{6b}
\end{equation}
\end{widetext}
where $q_{1}, q_{2}$ denote the two independent loop momenta. Furthermore, $M^{(Feynman)}$ denotes the sum of the contributions obtained from the first term in the photon propagator \eqref{6a} and is, therefore, independent of the gauge parameters $n^{\mu}, \beta$. However, the second term in \eqref{6b} is manifestly gauge dependent. Since the factor $(p\sl - m)$ does not commute with $n\sl$, it cannot be moved to one of the ends to annihilate the spinor. This, of course, implies that $\widetilde{M} = \widetilde{M} (n)$ if \eqref{6} holds.  The reason for this unexpected behavior can be traced to the fact that in non-covariant gauges with an additional structure, $\overline{u} S^{-1} (p)u\big|=0$ does not imply that $S^{-1} (p)u\big|=0$ so that $\widetilde{M}$ determined by relation \eqref{6} does not correspond to the pole of the propagator beyond the lowest order. The difference between $\widetilde{M}$ in \eqref{6} and the pole of the propagator can be algebraically seen to start at two loops (and beyond) \cite{new}. 

Let us consider massive QED in $n$ space-time dimensions (everything we say generalizes to QCD in a straightforward manner \cite{new}) described by the Lagrangian density
\begin{equation}
{\cal L}_{\rm inv} = - \frac{1}{4}\, F_{\mu\nu}F^{\mu\nu} + \overline{\psi} (i D\!\sl - m) \psi.\label{7}
\end{equation}
To study the question of gauge independence of the pole of the fermion propagator, covering both aspects of possible gauge dependence as discussed earlier, we choose a gauge fixing Lagrangian density given by \cite{taylor}
\begin{equation}
{\cal L}_{\rm\sc GF} = - \frac{1}{2}\left(\Lambda^{\mu} (\partial) A_{\mu}\right)^{2},\label{8}
\end{equation}
where
\begin{equation}
\Lambda^{\mu} (\partial) = \alpha \partial^{\mu} + \beta \partial^{\mu}_{\sc L},\quad \partial^{\mu}_{\sc L} = \frac{(n\cdot \partial)}{n^{2}}\,n^{\mu},\label{9}
\end{equation}
and $\alpha,\beta$ are arbitrary constant parameters. Clearly, when $\beta = 0$, the gauge fixing \eqref{8} corresponds to the class of covariant gauges; when $\alpha=0$ equation \eqref{8} leads to the class of generalized axial gauges and when $\beta = - \alpha$ and $n^{2} >0$, the gauge fixing Lagrangian density \eqref{8} corresponds to the class of generalized Coulomb gauges. Therefore, this general class of gauges interpolate between the covariant, the axial and the Coulomb gauges. The gauge fixing Lagrangian density depends on three independent parameters which we compactly denote as 
\begin{equation}
\phi_{(a)} = (\alpha, \beta, n^{\mu}).\label{9a}
\end{equation} 
Furthermore, for the purposes of manifest BRST invariance (which is essential for deriving the Nielsen identities), we note that the gauge fixing Lagrangian density can be written with an auxiliary field as
\begin{equation}
{\cal L}_{\sc\rm GF} = \frac{1}{2}\, F^{2} + \left(\Lambda^{\mu} (\partial) F\right) A_{\mu}.\label{10}
\end{equation}
The ghost Lagrangian density corresponding to this general class of gauge choice is given by
\begin{equation}
{\cal L}_{\rm ghost} = \left(\Lambda^{\mu} (\partial) \overline{c}\right) \partial_{\mu}c,\label{11}
\end{equation}
and the combined Lagrangian density ${\cal L}_{\rm inv} + {\cal L}_{\rm\sc GF} + {\cal L}_{\rm ghost}$ is invariant under the standard BRST transformations of QED \cite{brst,das},
\begin{align}
& \delta A_{\mu} = \omega \partial_{\mu}c, &  & \delta F = 0,\notag\\
& \delta \psi = - ie\omega c \psi, &  & \delta \overline{\psi} = -ie\omega \overline{\psi} c,\notag\\
& \delta c = 0, & & \delta \overline{c} = - \omega F,\label{12}
\end{align}
where $\omega$ represents an arbitrary constant Grassmann parameter.

To derive the Green's functions of the theory and to determine their gauge variations (Nielsen identities), we introduce the source Lagrangian density
\begin{align}
{\cal L}_{\rm source} & = J^{\mu}A_{\mu} + JF + i\left(\overline{\chi}\psi - \overline{\psi}\chi\right) + i\left(\overline{\eta} c - \overline{c}\eta\right)\notag\\
& + ie\left(\overline{M} c\psi - \overline{\psi}c M\right) + \left(H_{(\alpha)} (\partial^{\mu}\overline{c}) + H_{(\beta)} (\partial^{\mu}_{\sc L}\overline{c})\right)A_{\mu}\notag\\
& + \beta H_{(n)\,\mu} (N^{\mu\nu}\overline{c})A_{\nu},\label{13}
\end{align}
where we have identified
\begin{equation}
N^{\mu\nu} = \frac{\partial \partial^{\nu}_{\sc L}}{\partial n_{\mu}} = \frac{(n\cdot \partial)}{n^{2}} \left(\eta^{\mu\nu} + \frac{\partial^{\mu}n^{\nu}}{n\cdot\partial} - \frac{2n^{\mu}n^{\nu}}{n^{2}}\right).\label{14}
\end{equation}
Therefore, the total Lagrangian density for the gauge fixed theory is given by
\begin{equation}
{\cal L}_{\rm\sc TOT} = {\cal L}_{\rm inv} + {\cal L}_{\rm\sc GF} + {\cal L}_{\rm ghost} + {\cal L}_{\rm source}.\label{15}
\end{equation}
We note that, in addition to the standard sources for the dynamical fields of the theory, we have introduced the sources $(\overline{M}, M)$ which lead respectively to the composite BRST variations of $(\psi, \overline{\psi})$ in \eqref{12}. We have also introduced three other sources
\begin{equation}
H_{(a)} = \left(H_{(\alpha)}, H_{(\beta)}, H_{(n)}^{\mu}\right),\label{16}
\end{equation}
such that the BRST variations of these three source terms lead to the gauge variations of ${\cal L}_{\rm inv} + {\cal L}_{\rm\sc GF} + {\cal L}_{\rm ghost}$ with respect to the three parameters $\phi_{(a)}$ defined in \eqref{9a}. 

If we make a field redefinition corresponding to the BRST transformations \eqref{12}, namely, $\varphi \rightarrow \varphi + \delta\varphi$ ($\varphi$ denotes generically all the field variables) inside the path integral, the generating functional would not change since the path integral involves integration over all field configurations (the generating functional depends only on the sources and not on the fields). This leads to a Master identity for the generating functional for connected Green's functions. If we make a Legendre transformation of this with respect to the standard sources for the dynamical fields of the theory, we obtain the Master identity in terms of the effective action of the theory (Nielsen identity). This identity describes how the effective action changes as the three parameters are individually varied. Taking various field derivatives of this identity and setting all the fields to zero, one can determine how any 1PI amplitude changes with a change in the gauge fixing parameters. For example, by taking the functional derivative with respect to $\frac{\delta^{2}}{\delta\psi_{\beta} (p) \delta \overline{\psi}_{\alpha}(-p)}$ and setting all fields to zero we obtain \cite{schubert}
\begin{equation}
\frac{\partial S^{-1}_{\alpha\beta} (p)}{\partial \phi_{(a)}} = {\cal F}^{(a)}_{\alpha\gamma} (p) S^{-1}_{\gamma\beta} (p) + S^{-1}_{\alpha\gamma} (p) {\cal G}^{(a)}_{\gamma\beta} (p),\label{18}
\end{equation}
where we have identified the three point amplitudes  
\begin{align}
{\cal F}^{(a)}_{\alpha\beta} (p) & = \frac{\delta^{3}\Gamma}{\delta\overline{\psi}_{\alpha} (-p)\delta H_{(a)} (0)\delta M_{\beta}(p)}\Big|,\notag\\
{\cal G}^{(a)}_{\alpha\beta} (p) & = \frac{\delta^{3}\Gamma}{\delta\overline{M}_{\alpha}(-p)\delta H_{(a)} (0) \delta\psi_{\beta} (p)}\Big|,\label{19}
\end{align}
with the restriction implying setting of all fields to zero after taking the functional derivatives. Equation \eqref{18} shows how the fermion two point function changes with respect to the three parameters $(\alpha,\beta,n^{\mu})$. We note that since there are no vertices corresponding to ${\cal F}^{(a)}_{\alpha\beta} (p), {\cal G}^{(a)}_{\alpha\beta} (p)$ at the tree level Lagrangian density in \eqref{15}, these amplitudes are nontrivial only at one loop and beyond. Therefore, the Nielsen identity \eqref{18} shows that the dependence of the fermion two point function on the gauge fixing parameter arises only at one loop and beyond.

As we have mentioned earlier, the final ingredient in the proof of the gauge independence of the pole of the fermion propagator is to relate the denominator of the propagator to the two point function. We note from \eqref{4} that we can parameterize the fermion two point function to all orders as
\begin{align}
S^{-1} (p) & = (1-B) p\sl - m(1+A) -Cp\sl_{\sc L} \notag\\
&\qquad - \frac{mD}{p_{\sc L}^{2}} (p\sl_{\sc L}p\sl - p\sl p\sl_{\sc L}),\label{20}
\end{align}
Here the dimensionless coefficients $A,B,C,D$, of course, depend on $p^{2},p_{\sc L}^{2}, m$, but also on the gauge fixing parameters $n^{2},\alpha,\beta$. The fermion propagator can now be determined to have the form
\begin{equation}
S (p) = \frac{\cal N}{\cal D} = - \frac{{\cal C} (S^{-1} (p))^{T} {\cal C}^{-1}}{\cal D},\label{21}
\end{equation}
where ${\cal C}$ denotes the charge conjugation matrix, $T$ the matrix transpose and the scalar denominator has the explicit form
\begin{align}
{\cal D} & = \left((1-B)^{2} - \frac{m^{2}D^{2}}{p_{\sc L}^{2}}\right)p^{2} - m^{2} (1+A)^{2} \notag\\
&\quad + \left(C^{2} - 2(1-B)C + \frac{m^{2}D^{2}}{p_{\sc L}^{2}}\right)p_{\sc L}^{2}.\label{22}
\end{align}
This shows that the propagator is manifestly gauge dependent.The pole of the propagator, of course,  corresponds to the zero of the denominator ${\cal D}$. We note from \eqref{21} that we can write 
\begin{align}
{\cal D} \mathbbm{1} & = - S^{-1} (p) {\cal C} (S^{-1}(p))^{T} {\cal C}^{-1}\notag\\
 & = - {\cal C} (S^{-1} (p))^{T} {\cal C}^{-1} S^{-1}(p),\label{23}
\end{align}
where $\mathbbm{1}$ denotes the identity matrix in the spinor space. In $n$ dimensions this corresponds to the $2^{[n/2]}\times 2^{[n/2]}$ unit matrix where $[n/2]$ denotes the integer part of $n/2$. Taking the trace of \eqref{23} we obtain
\begin{equation}
{\cal D} = - \frac{1}{2^{[n/2]}}\, {\rm Tr} \left(S^{-1}(p) {\cal C} (S^{-1}(p))^{T} {\cal C}^{-1}\right).\label{24}
\end{equation}

We can now study the gauge parameter dependence of the denominator of the propagator \eqref{24} using the Nielsen identity \eqref{18}. Using the relations \eqref{23} and \eqref{24} as well as the cyclicity of trace, we obtain
\begin{align}
\frac{\partial {\cal D}}{\partial \phi_{(a)}} & =  {\cal D}\,{\rm Tr}\left({\cal F}^{(a)}+{\cal G}^{(a)} + {\cal C} ({\cal F}^{(a)} +{\cal G}^{(a)})^{T} {\cal C}^{-1}\right)\notag\\
& = 2 {\cal D}\,{\rm Tr}\left({\cal F}^{(a)}+{\cal G}^{(a)}\right) .\label{25}
\end{align}
This shows that the denominator indeed is gauge dependent as we have pointed out earlier. However, we note that, if the amplitudes $({\cal F}^{(a)} (p), {\cal G}^{(a)}(p))$ are well behaved, the zero of the denominator ${\cal D}$ (which corresponds to the pole of the propagator) is, in fact, gauge parameter independent, namely,
\begin{equation}
\frac{\partial {\cal D}}{\partial \phi_{(a)}}\Big|_{{\cal D}=0} = 0.\label{26}
\end{equation}
This would, in fact, be the case when the theory does not have any infrared divergences or mass shell singularities at the pole of the propagator. Actually, this turns out to be the case in perturbative QED (and in QCD) in $4$ space-time dimensions \cite{kronfeld}. On the other hand, in lower dimensions, $n < 4$, such singularities can be present in the amplitudes $({\cal F}^{(a)} (p), {\cal G}^{(a)} (p))$ so that the pole of the propagator may become gauge dependent. This was explicitly studied in the massive Schwinger model (massive QED in $1+1$ dimensions) in \cite{schubert}. Noting that near the zero of ${\cal D}$ (or the pole of the propagator), we can write
\begin{equation}
{\cal D}\xrightarrow{p^{2}\rightarrow M^{2}_{\rm phys}} {\cal B} (p^{2} - M^{2}_{\rm phys}),\label{27}
\end{equation}
where the coefficient ${\cal B}$ is related to the wave function normalization $Z_{2}^{-1}$, equation \eqref{26} leads to
\begin{equation}
\frac{\partial M_{\rm phys}}{\partial \phi_{(a)}} = 0.\label{28}
\end{equation}
Namely, the pole of the propagator or the physical mass of the fermion is independent of the three gauge fixing parameters $(\alpha, \beta, n^{\mu})$. Furthermore, since the covariant, the axial and the Coulomb gauges correspond to specific values of these parameters which $M_{\rm phys}$ is independent of, the physical mass (or the location of the pole of the propagator) is the same in all three gauges. It is worth emphasizing that this direct and simple demonstration of gauge invariance of the pole of the fermion propagator involves only three basic elements: choice of an interpolating gauge, the Nielsen identity for the gauge variation of the fermion two point function and the relation of this function to the denominator of the propagator.

To conclude, in this letter we have chosen a general class of gauge fixing terms to study the question of both aspects of gauge independence of the pole of the fermion propagator in a gauge theory. This gauge fixing interpolates between the covariant, the axial and the Coulomb gauges. We have derived the Nielsen identity for the gauge variation of the fermion two point function in this gauge. Using this we have shown that the pole of the propagator is not only independent of the gauge fixing parameters, but is also the same in the covariant, the axial and the Coulomb gauges. Details of our derivations as well as a discussion of other relevant aspects of this problem will be presented separately \cite{new}.

\bigskip

\noindent{\bf Acknowledgments}
\medskip

 A. D. would like to thank the Departamento de F\'{i}sica Matematica for hospitality where this work was done. This work was supported in part by USP and by CNPq.

\end{document}